\documentclass[referee]{aa} 
\usepackage{graphicx}
\usepackage{natbib}
\begin{document}
\titlerunning{Convection with RSST}
   \title{Numerical Calculation of Convection with Reduced Speed of
   Sound Technique}

   \author{H. Hotta
          \inst{1}
          \and
          M. Rempel\inst{2}
	  \and
	  T. Yokoyama\inst{1}
	  \and
	  Y. Iida\inst{1}
	  \and
	  Y. Fan\inst{2}
          }
	  
   	  \institute{
   	  Department of Earth and Planetary Science, University of
   	  Tokyo, 7-3-1 Hongo, Bunkyo-ku, Tokyo 113-0033, Japan
          \and
	  High Altitude Observatory, National Center for Atmospheric Research,
	  Boulder, CO, USA}


 
  \abstract
  {The anelastic approximation is often  adopted in numerical
  calculation with low Mach number, such as stellar internal
  convection. This approximation requires frequent global
  communication, because of an elliptic partial differential
  equation. Frequent global communication is negative factor for the
  parallel computing with a large number of CPUs.}
  {The main purpose of this paper is to test the validity of a
  method that artificially reduces the speed of sound for the
  compressible fluid equations
  in the context
  of stellar internal convection. The reduction of speed of sound allows
  for larger time steps in spite of low Mach number, while the numerical
  scheme remains fully explicit and the mathematical system is
  hyperbolic and thus does not require frequent global
  communication.}
  {Two and three dimensional compressible hydrodynamic equations are solved
  numerically. 
  Some statistical quantities of solutions computed with
  different effective Mach numbers (due to reduction of speed of sound)
  are compared to test the validity of our approach.}
  {Numerical simulations with artificially reduced speed of sound
  are a valid approach as long as the effective Mach number (based on the
  reduced speed of sound) remains less than 0.7.}
  {}


   \maketitle

\section{Introduction}\label{introduction}
Turbulent thermal convection in the solar convection zone plays a key role for the
maintenance of large scale flows (differential rotation, meridional
flow) and solar magnetic activity.
The angular momentum transport of convection maintains the global
mean flows. Global
flows relate to the generation of global magnetic field, i.e. the solar
dynamo.
Differential
rotation bends the pre-existing poloidal field and generates the strong
toroidal field ($\Omega$ effect) and mean meridional flow
transports the magnetic flux equatorward at the base of the convection zone
\citep{1995A&A...303L..29C,1999ApJ...518..508D}.
The internal structures of the solar differential rotation and the
meridional flow are revealed by the
helioseismology \citep[see review by][]{2003ARA&A..41..599T}.
Some mean field studies have reproduced these global flow
\citep{1993A&A...276...96K,2001A&A...366..668K,2005ApJ...622.1320R,2011ApJ...740...12H}. 
These studies, however, used some kinds of parameterization of the turbulent
convection, i.e. turbulent viscosity and turbulent angular momentum
transport. Thus, a self-consistent thorough understanding of global structure
requires the detailed investigation of the turbulent thermal convection.
Turbulent convection is important also in the magnetic field itself. The
strength of the next solar maximum in the prediction by the mean field model 
significantly depends on the turbulent diffusivity
\citep{2006ApJ...649..498D,2007PhRvL..98m1103C,2008ApJ...673..544Y}.
In addition, the turbulent diffusion has an important role in the parity of
solar global field, the strength of polar field and so on
\citep{2010ApJ...709.1009H,2010ApJ...714L.308H}.\par
There are already numerous LES numerical simulations on the solar and
stellar convection
\citep{1977GApFD...8...93G,1981ApJS...46..211G,1984JCoPh..55..461G,2000ApJ...532..593M,2006ApJ...641..618M,2008ApJ...689.1354B}
and these magnetic fields
\citep{1981ApJS...46..211G,2004ApJ...614.1073B,2010ApJ...711..424B,2011ApJ...731...69B}.
In these studies the anelastic approximation is adopted to avoid the
difficulty which is caused by the high speed of sound. At the base of
convection zone the speed of sound is about $200\ \mathrm{km\ s^{-1}}$.
In contrast the speed of convection is thought to be 50
$\mathrm{m\ s^{-1}}$ \citep{2004suin.book.....S}, so the time step must
be shortened due to the CFL condition in explicit fully compressible method, even when we
are interested in phenomena related to convection. In the anelastic
approximation, equation of continuity is treated as
\begin{eqnarray}
 \nabla\cdot(\rho_0 {\bf v})=0,
\end{eqnarray}
where $\rho_0$ is stratified background density and ${\bf v}$ denotes the
velocity. 
The anelastic approximation assumes that the speed of sound is
essentially infinite, resulting in an instantaneous adjustment of
pressure to flow changes. This is achieved by solving an elliptic
equation for the pressure, which filters out the propagation of
sound waves. As a result the time step is only limited by the much
lower flow velocity. However, due to the existence of an elliptic
the anelastic approximation has a weak point.
The numerical calculation
in parallel computing
requires the frequent global communication.
At the present time the efficiency of scaling in parallel computing is
saturated with about 2000-3000 CPUs in solar global simulation with
pseudo-spectral method
(M. Miesch private communication).
More and more resolution, however, is thought to be needed to understand the precise
mechanism of the angular momentum and energy transport by the turbulent
convection and the behavior of magnetic field especially in thin
magnetic flux tube.\par
In this paper, we test the validity of a different approach to
circumvent the severe numerical time step constraints in low
Mach number flows. We use a method in which the speed of sound is
reduced artificially by transforming the equation of continuity
to \cite[see, e.g.,][]{2005ApJ...622.1320R}
\begin{eqnarray}
 \frac{\partial \rho}{\partial t}=-\frac{1}{\xi^2}\nabla\cdot(\rho {\bf v}),\label{RSST}
\end{eqnarray}
where $t$ denotes the time. Using this equation, the effective speed of
sound becomes $\xi$ times smaller, but otherwise the dispersion
relationship
for sound wave
remains unchanged (wave speed is dropped for all wavelength equally).
Since this technique does not change the hyperbolic
character of the underlying equations, the numerical treatment can
remain fully explicit and thus does not require global communication
in parallel computing. We will call the technique in the following
the Reduced Speed of Sound Technique (RSST).
This technique has been used previously by Rempel (2005, 2006) in
mean field models for solar differential rotation and non-kinematic
dynamos, which essentially solve the full set of time dependent
axisymmetric MHD equations. Those solutions were however
restricted to the relaxation toward a stationary state or very slowly
varying problems on the time scale of the solar cycle. Here we will
apply this approach to thermal convection, where the intrinsic time
scales are substantially shorter. To this end we study two and three
dimensional convection, in particular the latter will be
non-stationary and turbulent.
\par
The detailed setting of test calculation is given in Section \ref{model}.
The results of our calculations are given in Section \ref{result}.
We summarize our paper and give discussion of the RSST in Section \ref{summary}.
\section{Model}\label{model}
\subsection{Equations}
The two or three dimensional equation of continuity,
equation of motion, equation of energy, equation of state, are solved in Cartesian
coordinate $(x,z)$ or $(x,y,z)$, where $x$ and $y$ denote the horizontal
direction and $z$ denotes the vertical direction.
The basic assumptions underlying this study are as follows.
\begin{enumerate}
 \item Time independent hydrostatic reference state.
 \item The perturbations caused by thermal convection are small, i.e.
$\rho_1\ll\rho_0$ and $p_1\ll p_0$, Here $\rho_0$ and $p_0$ denote the
       reference state values, whereas $\rho_1$ and $p_1$ are the
       fluctuations of density and pressure, respectively.
       Thus a linearized equation of state is used as eq. (\ref{e:e4})
 \item The profile of the reference entropy $s_0(z)$ is a
steady state solution of the thermal diffusion equation
$\nabla\cdot(K\rho_0T_0\nabla s_0)=0$ with constant $K$.
\end{enumerate}
The formulations are
almost same as \cite{2003ApJ...582.1206F}.
Equations are expressed as,
\begin{eqnarray}
 && \frac{\partial \rho_1}{\partial t}=-\frac{1}{\xi^2}
  \nabla\cdot(\rho_0{\bf v})\label{e:e1},\\
 && \frac{\partial {\bf v}}{\partial t}=-({\bf
  v}\cdot\nabla){\bf v}-\frac{\nabla p_1}{\rho_0}
  -\frac{\rho_1}{\rho_0}g{\bf e_z}+\frac{1}{\rho_0}\nabla\cdot{\bf \Pi},\label{e:e2}\\
 && \frac{\partial s_1}{\partial t}=-({\bf v}\cdot\nabla)(s_0+s_1)
  +\frac{1}{\rho_0T_0}\nabla\cdot(K\rho_0T_0\nabla s_1)\nonumber\\
  &&+\frac{\gamma-1}{p_0}({\bf
  \Pi}\cdot\nabla)\cdot{\bf v},\label{e:e3}\\
  && p_1 = p_0 \left(\gamma \frac{\rho_1}{\rho_0}+s_1\right),\label{e:e4}
\end{eqnarray}
where $T_0(z)$, and $s_0(z)$ denote reference
temperature, and entropy, respectively and ${\bf e_z}$
denotes the unit vector along the $z$-direction. $\gamma$ is the ratio
of specific heats, with the value for an ideal gas being $\gamma=5/3$.
$s_1$ denotes the fluctuation of
entropy from reference atmosphere. Note that the entropy
is normalized by specific heat
capacity at constant volume $c_\mathrm{v}$.
The quantity $g$
is the gravitational acceleration, which is assumed to be constant. The
quantity ${\bf \Pi}$ denotes the viscous stress tensor,
\begin{eqnarray}
 \Pi_{ij}=\rho_0\nu
\left[
\frac{\partial v_i}{\partial x_j}+\frac{\partial v_j}{\partial x_i}
-\frac{2}{3}(\nabla\cdot{\bf v})\delta_{ij}
\right],
\end{eqnarray}
and $\nu$ and $K$ denote the viscosity and thermal diffusivity,
respectively. $\nu$ and $K$ are assumed to be constant throughout the
simulation domain.
We assume for the reference atmosphere a weakly superadiabatically
stratified polytrope:
\begin{eqnarray}
&& \rho_0(z)=\rho_r
\left[
1-\frac{z}{(m+1)H_r}
\right]^m, \\
&& p_0(z)=p_r
\left[
1-\frac{z}{(m+1)H_r}
\right]^{m+1}, \\
&& T_0(z)=T_r
\left[
1-\frac{z}{(m+1)H_r}
\right],\\
&& H_p(z) = \frac{p_0}{\rho_0g},\\
&& \frac{ds_0}{dz}=-\frac{\gamma\delta(z)}{H_p(z)},\\
&&\delta(z)=\delta_r\frac{\rho_r}{\rho_0(z)},
\end{eqnarray}
where $\rho_\mathrm{r}$, $p_\mathrm{r}$, $T_\mathrm{r}$, $H_\mathrm{r}$,
and $\delta_\mathrm{r}$ denote the values of $\rho_0$, $p_0$, $T_0$,
$H_0$ (the pressure scale height) and $\delta$ (the non-dimensional
superadiabaticity) at the bottom boundary $z=0$. Since $|\delta|\ll 1$,
the value $m$ is nearly equal to the adiabatic value, meaning
$m=1/(\gamma-1)$.
The strength of
the diffusive parameter $\nu$ and $K$ is expressed with following
non-dimensional parameters: the Reynolds number
$\mathrm{Re}\equiv v_\mathrm{c}H_\mathrm{r}/\nu$, and the Prandtl number
$\mathrm{Pr}\equiv\nu/K$, where the velocity unit
$v_\mathrm{c}\equiv(8\delta_\mathrm{r}gH_\mathrm{r})^{1/2}$. Note that
in this paper the unit of time is $H_\mathrm{r}/v_\mathrm{c}$. In all
calculations, we set $\mathrm{Pr}=1$.
\subsection{Boundary Conditions and Numerical Method}
We solve equations (\ref{e:e1})-(\ref{e:e4}) numerically.
At the horizontal boundaries ($x=0,\ L_x$ and $y=0,\ L_y$), periodic
boundary conditions are adopted for all variables.
At the top and the bottom boundaries impenetrative and stress free
boundary conditions are adopted for the velocities and the entropy is fixed:
\begin{eqnarray}
 v_z = 0,\\
 \frac{\partial v_x}{\partial z}=0,\\
 \frac{\partial v_y}{\partial z}=0,\\
s_1 = 0.
\end{eqnarray}
At both top and bottom boundaries ($z=0$ and $z=L_z$), we set $p_1$ in
the ghost cells such
that the right hand side of the z-component of eq. (\ref{e:e2}) is zero
at the boundary (which is between ghost cells and domain cells), where
the ghost cells are the cells beyond the physical boundary. \par
We adopt the fourth-order space-centered difference for each derivative.
The first spatial derivatives of quantity $q$ is given by
\begin{eqnarray}
\left(
\frac{\partial q}{\partial x}
\right) _i
 = \frac{1}{12\Delta x}(-q_{i+2}+8q_{i+1}-8q_{i-1}+q_{i-2}),\label{forth}
\end{eqnarray}
where $i$ denotes the index of the grid position along a particular
spatial direction. The numerical solution of the system is advanced in
time with an explicit fourth-order Runge-Kutta scheme. The system of
partial equations can be written as
\begin{eqnarray}
 \frac{\partial {\bf U}}{\partial t}={\bf R(U)}
\end{eqnarray}
${\bf U}_{n+1}$, which is the value at $t_{n+1}=(n+1)\Delta t$ is calculated in
four steps:
\begin{eqnarray}
&& {\bf U}_{n+\frac{1}{4}}={\bf U}_{n}+\frac{\Delta t}{4}{\bf R}({\bf U}_{n}),\\
&& {\bf U}_{n+\frac{1}{3}}={\bf U}_{n}+\frac{\Delta t}{3}{\bf R}({\bf U}_{n+\frac{1}{4}}),\\
&& {\bf U}_{n+\frac{1}{2}}={\bf U}_{n}+\frac{\Delta t}{2}{\bf R}({\bf U}_{n+\frac{1}{3}}),\\
&& {\bf U}_{n+1}={\bf U}_{n}+\Delta t{\bf R}({\bf U}_{n+\frac{1}{2}}).
\end{eqnarray}
\par
The maximum allowed time step, $\Delta t_\mathrm{max}$, is determined by
the CFL criterion. When both advection and diffusion terms are included
in calculation, the time step reads,
\begin{eqnarray}
 \Delta t_\mathrm{max} = \min(\Delta t_\mathrm{ad}, \Delta t_\mathrm{v}).
\end{eqnarray}
Here
\begin{eqnarray}
 \Delta t_\mathrm{ad}=c_\mathrm{ad}\frac{\min(\Delta x, \Delta y, \Delta z)}{c_\mathrm{tot}}.
\end{eqnarray}
$c_\mathrm{tot}$ is the total wave speed:
\begin{eqnarray}
 c_\mathrm{tot} = |v|+c'_\mathrm{s},
\end{eqnarray}
where the effective speed of sound is expressed as:
\begin{eqnarray}
c'_\mathrm{s}=\frac{1}{\xi}\sqrt{\gamma\frac{p_0}{\rho_0}}.
\end{eqnarray}
 The time step determined by diffusion term is
\begin{eqnarray}
 \Delta t_\mathrm{v} = 
c_\mathrm{v}\frac{\min(\Delta x^2, \Delta y^2, \Delta z^2)}{\max(K,\nu)}.
\end{eqnarray}
$c_\mathrm{ad}$ and $c_\mathrm{v}$ are safety factors of order unity.
 
Using $\delta_\mathrm{r}=1\times10^{-4}$, the original speed of
sound is about
35$v_\mathrm{c}$ at the bottom and
12$v_\mathrm{c}$ at the top boundary, respectively. In all calculation
$\Delta x\sim2.3\times10^{-2}H_\mathrm{r}$. Thus if we use $c_\mathrm{ad}=1$
and $c_\mathrm{v}=1$, $\Delta t_\mathrm{ad}=6.4\times10^{-4}\ H_\mathrm{r}/v_\mathrm{c}$ and
$\Delta t_\mathrm{v}=1.5\times10^{-1}\ H_\mathrm{r}/v_\mathrm{c}$.
For all the $\xi$ values considered in this paper the time step remains
restricted by the (reduced) speed of sound, 
thus the calculation has about $\xi$ times better
efficiency with the RSST.
\section{Result}\label{result}
\subsection{Two Dimensional Study}
\begin{figure}[htbp]
 \centering
 \includegraphics[width=15cm]{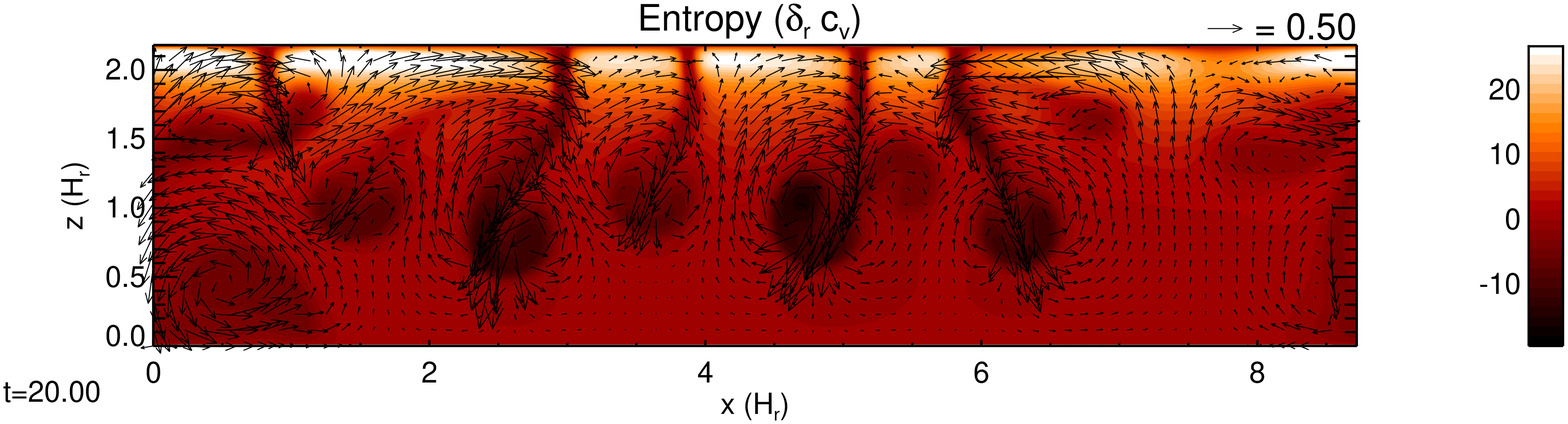}
 \includegraphics[width=15cm]{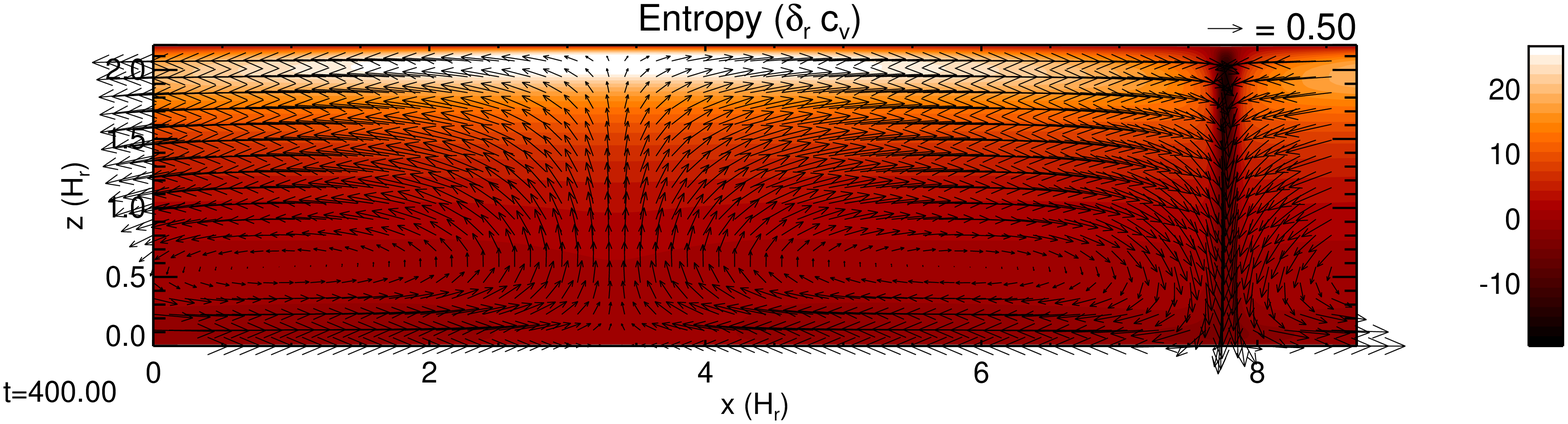}
\caption{Time-development of entropy in a two-dimensional calculation
 with parameters of case 1. (Top panel) $t=20$. (Bottom panel) $t=400$.\label{2d_conv}}
\end{figure}

\begin{figure}[htbp]
 \includegraphics[width=15cm]{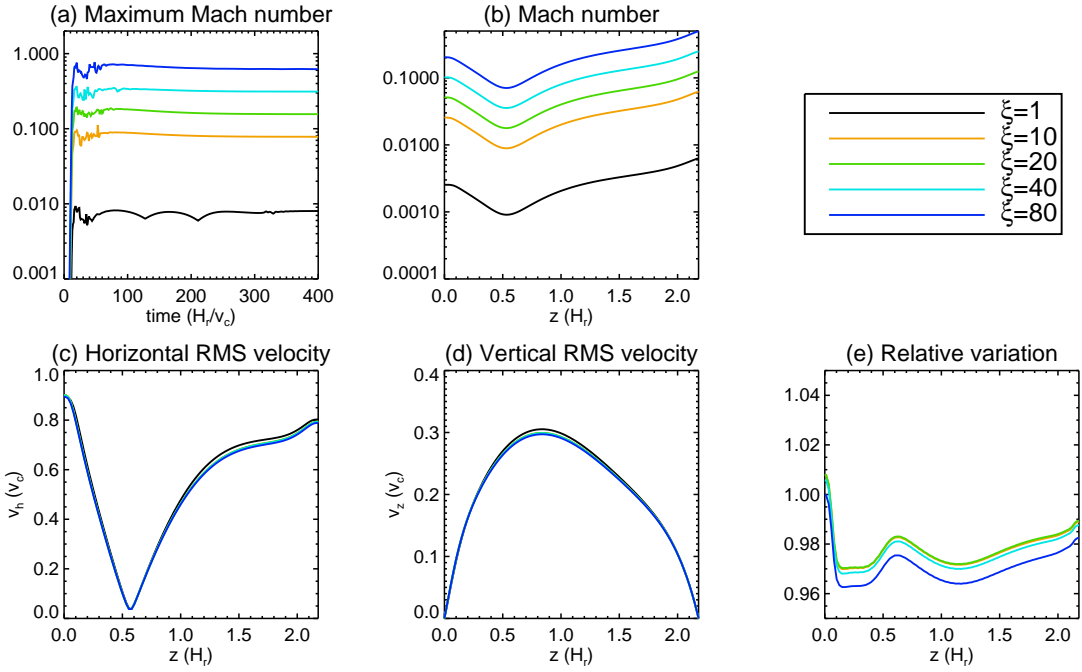}
\caption{Some quantities in two dimensional calculation.(a) Maximum Mach
 number of each time step. (b) Distribution
 of Mach number estimated with the RMS velocity. (c) Distribution of
 horizontal RMS velocity $v_\mathrm{h}$. (d) Distribution of vertical
 RMS velocity. (e) Ratio between the RMS velocities with each $\xi$
 and $\xi=1$. \label{2d_compare}}
\end{figure}

Four two-dimensional calculations with the RSST and one calculation
without approximation are carried out (case 1-5). The
values of free-parameters are given in Table \ref{param}.
The superadiabaticity ($\delta$) is $1\times10^{-6}$ at
the bottom and about $2\times 10^{-5}$ at the top boundary. It is almost the same
superadiabaticity value as the base of solar convection zone.
Fig. \ref{2d_conv} shows the time-development of entropy. In the
beginning the non-linear time dependent convection developed (top
panel), which transitions to a steady state at later times
(bottom panel). In the steady state, we
compare the RMS velocity with different $\xi$. The RMS velocity is
defined as:
\begin{eqnarray}
 v_\mathrm{RMS}=\sqrt{\frac{1}{L_xL_y}\int_0^{L_x}\int_0^{L_y}v^2dxdy}.
\end{eqnarray}
Fig. \ref{2d_compare} shows our results with the value of $\xi$ being
from 1 to 80. The effective Mach number is
defined as
$M_\mathrm{A}=v_\mathrm{RMS}/c'_\mathrm{s}$.
 If we have larger $\xi$ value than 80, we cannot obtain
stationary state, since there are some shock generated by supersonic
convection. The discussion of unsteady convection is given in
the next session of three-dimensional calculations.
Even though the Mach number reaches 0.6 using $\xi=80$ (panel a),
the horizontal and vertical RMS velocity is almost same as those
calculated with $\xi=1$ (without RSST). The ratio between the RMS
velocities with each $\xi$ and $\xi=1$ are shown in
Fig. \ref{2d_compare}e. The deviation is always a few percent.
 This result is not
surprising since in the stationary state the equation of continuity
becomes
\begin{eqnarray}
 0=\frac{1}{\xi^2}\nabla\cdot(\rho_0 {\bf v}),
\end{eqnarray}
whose solution must not depend on the value of $\xi$.
We note that the cell size is to some degree affected by the aspect
ratio of the domain. This does not affect our conclusions since this
influence is the same for all values of $\xi$ considered. To confirm the
robustness of our conclusions we repeated this experiment with a
wider domain ($26.16H_r\times2.18H_r$ instead of $8.72H_r\times2.18H_r$) in which we find
4 steady convection cells with about $10\%$ different RMS velocities. Also
here we find a dependence on $\xi$ similar to that shown in Fig. \ref{2d_compare}, i.e.
the solutions show differences of only a few percent as long as $\xi<80$.
%
In this section we confirm that the RSST is valid for the
two-dimensional stationary convection when the effective Mach number is
less than 1, i.e. $\xi<80$ with $\delta_\mathrm{r}=1\times10^{-6}$. If
we use $\delta_\mathrm{r}=1\times10^{-4}$ (result is not shown), the
criterion becomes $\xi<8$ in two dimensional calculation.
\par
\begin{figure}[htbp]
\includegraphics[width=10cm]{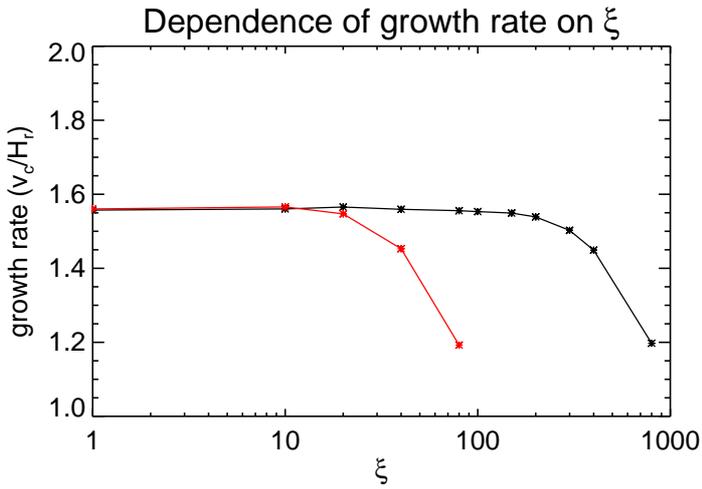}
\caption{Behavior in linear phase. Dependence of growth rate on $\xi$ is
 shown. Black and red lines show the results with
 $\delta_\mathrm{r}=1\times10^{-6}$ and $1\times10^{-4}$, respectively.\label{linear}}
\end{figure}

As a next step we investigate the dependence of the linear growth
rate on $\xi$ during the initial (time dependent) relaxation phase
toward the final stationary state.
Fig. \ref{linear} shows the linear growth of maximum perturbation
density $\rho_1$ with different $\xi$. 
Black and red lines show the results with
$\delta_\mathrm{r}=1\times10^{-6}$ and
$\delta_\mathrm{r}=1\times10^{-4}$, respectively.
These calculation
parameters are not given in Table \ref{param}. 
In the calculation with $\delta=1\times10^{-6}$ the growth rate decreases for values
of $\xi >300$,
 whereas it occurs $\xi> 30$ in the calculation with
$\delta_\mathrm{r}=1\times10^{-4}$.
The reason can be explained as follows: In the convective
instability, upflow (downflow) generates
positive (negative) entropy perturbation and then negative
(positive) density perturbation is generated by the sound wave. If the
speed of sound is fairly slow, the generation mechanism of density
perturbation is ineffective. In the calculation with
$\delta_\mathrm{r}=1\times10^{-6}$ ($\delta_\mathrm{r}=1\times10^{-4}$),
the growth rate with $\xi=200$ ($\xi=20$), however, is almost same as that with
$\xi=1$, even though flow with $\xi=200$ ($\xi=20$) is expected to be the Mach
number $Ma=1.3$, i.e. supersonic convection flow
in the saturated state.\par

\subsection{Three Dimensional Study}\label{three}
\begin{figure}[htbp]
\includegraphics[width=15cm]{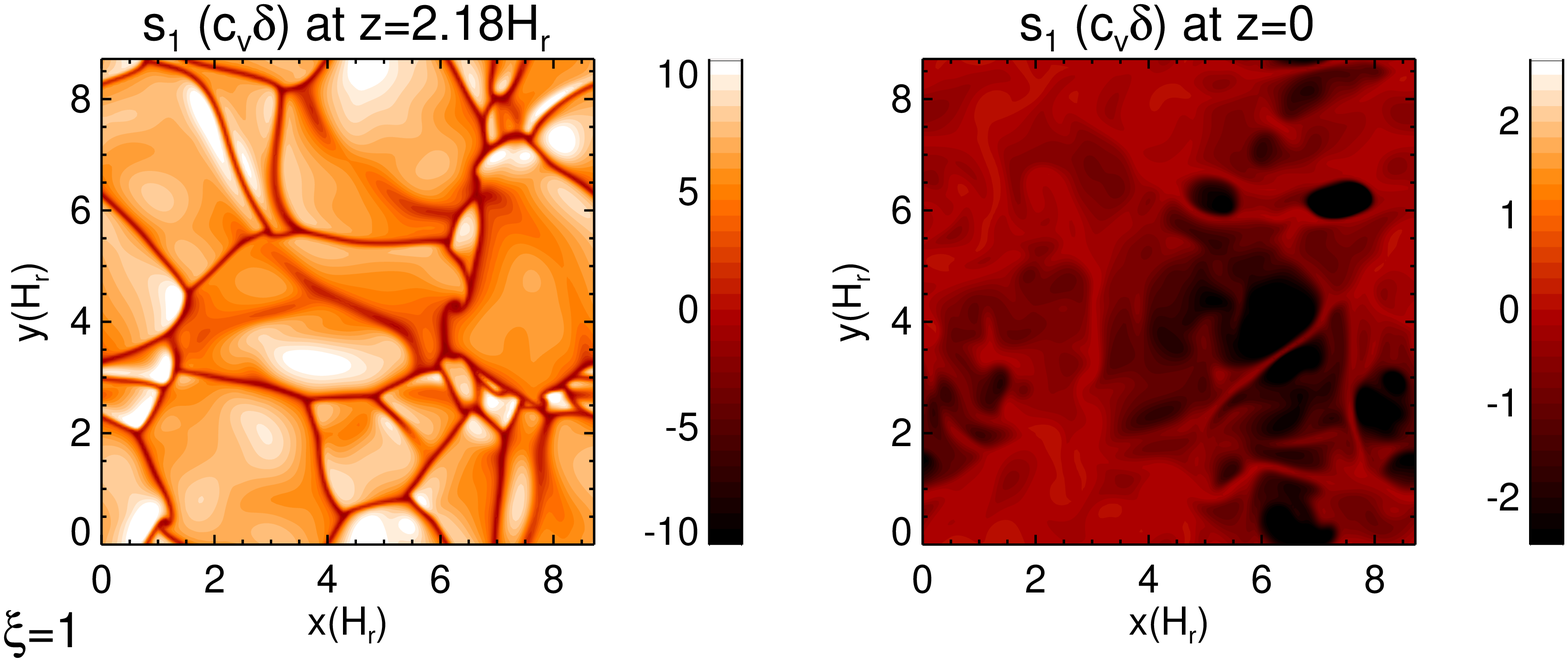}
\includegraphics[width=15cm]{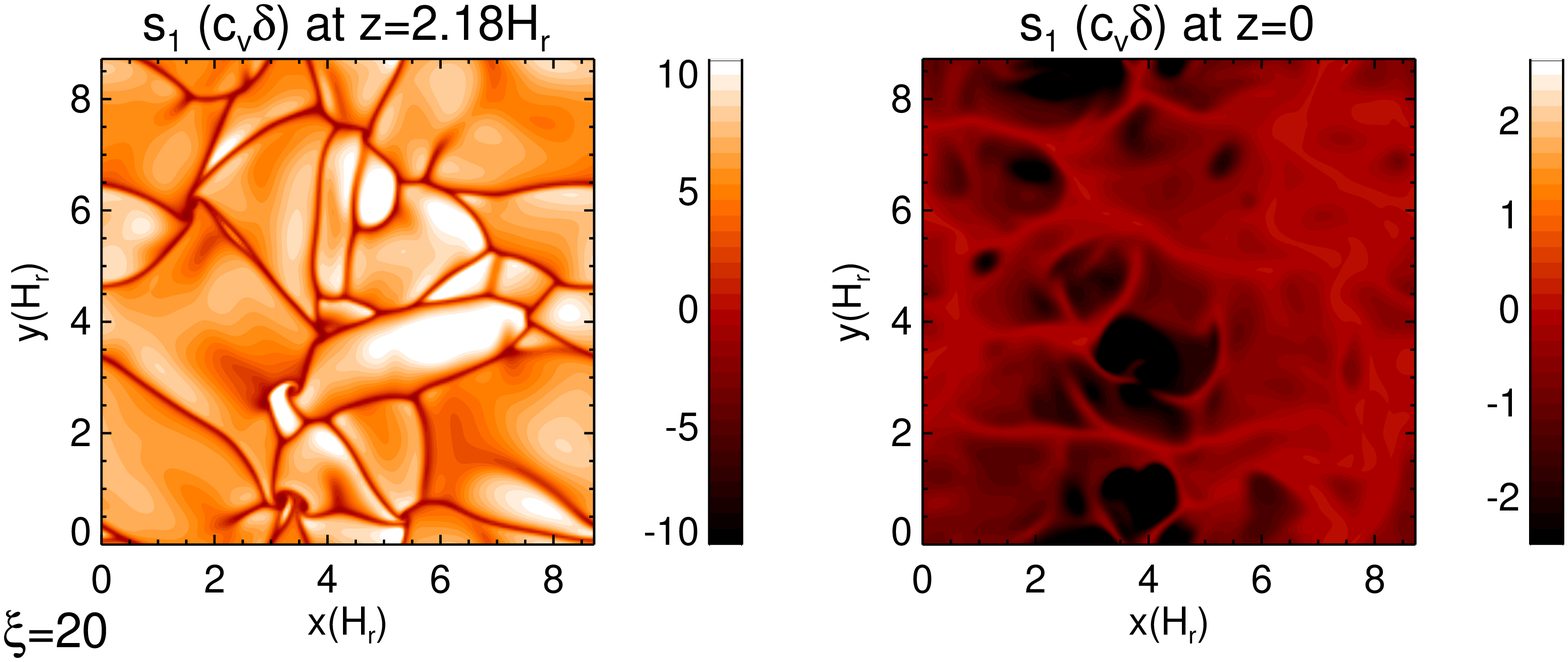}
\includegraphics[width=15cm]{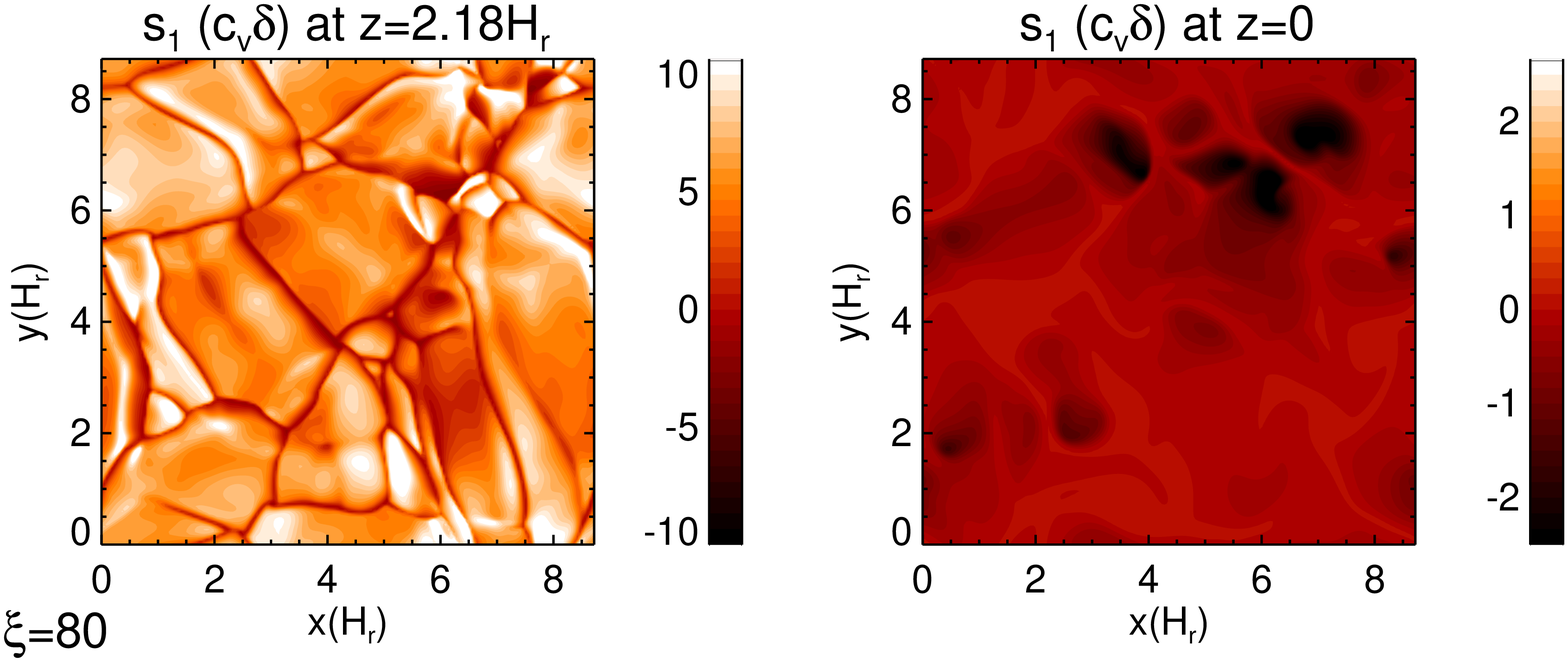}
\caption{Snapshots of entropy of the three-dimensional convection. Top,
 middle, and bottom panels correspond to $\xi=1$, 10 and 80
 respectively. Left (right) panels show entropy at top (bottom)
 boundary. (Animation is provided, 
 and  the difference between  $\xi=1$ and $80$ is best visible in
 the animation of Fig. \ref{3d_conv} that is provided with the online version.)
 \label{3d_conv}}
\end{figure}

In this section, we investigate the validity of the RSST with
three-dimensional unsteady thermal convections (case 6-13).
The value of superadiabaticity at the bottom boundary is $1\times10^{-4}$ and at
the top $2\times 10^{-3}$. Although this value is relatively large compared with
solar value, the expected speed of convection is much
smaller than speed of sound, so it is small enough to investigate the
validity of the RSST. 
Entropy of three-dimensional convections with $\xi=1$, 20, and 80 at
$ t=100H_\mathrm{r}/v_\mathrm{c}$ are
shown in Fig. \ref{3d_conv}. The convection is completely unsteady
and turbulent (animation is provided).
The appearance of convection with $\xi=20$
is almost same as that with $\xi=1$. This will be verified below by the
Fourier transformation and auto-detection technique.
On the other hand, the appearance of convection with $\xi=80$ (bottom panel)
is completely different from the others. 
 This difference is best visible in the animation of Fig. \ref{3d_conv} that is
provided with the online version.\par
\begin{figure}[htbp]
 \includegraphics[width=15cm]{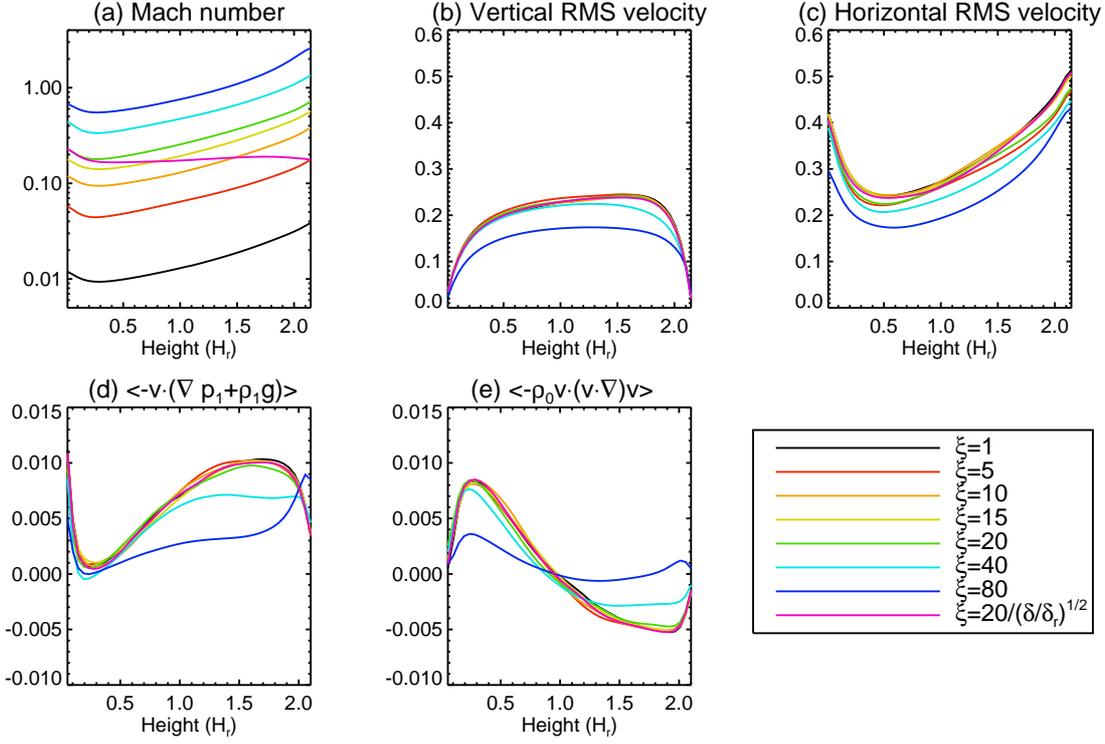}
\caption{ Some quantities averaged in time between $t=100$ and
 $200H_\mathrm{r}/v_\mathrm{c}$ in three dimensional calculation.(a)  Distribution
 of Mach number estimated with the RMS velocity. (b) Distribution of
 vertical RMS velocity $v_\mathrm{h}$. (c) Distribution of horizontal
 RMS velocity. (d) Distribution of RMS power density of pressure and
 buoyancy. (e) Distribution of RMS power density of inertia.\label{3d_compare}}
\end{figure}

RMS velocities with different $\xi$ are estimated as average
of values
between $t=100$ to $ 200H_\mathrm{r}/v_\mathrm{c}$
(Fig. \ref{3d_compare}: panel b and c). 
 Without the RSST, i.e. $\xi=1$ and using
$\delta_\mathrm{r}=1\times10^{-4}$, the Mach number is $1\times10^{-2}$
at the bottom and $4\times10^{-2}$ at the top boundary.
The RMS velocities at $\xi=40$ and 80 differ from those at the
 $\xi=1$ by less than $15\ \%$ and $30\ \%$, respectively.
When
we adopt $\xi=40$ and 80, the Mach number estimated by RMS velocity
exceeds unity (Fig. \ref{3d_compare}: panel a),
i.e. supersonic convection. This supersonic downflow frequently
generates shocks and positive entropy perturbation, thus
downflow is slowed. This is the reason why the RMS velocities with large
$\xi (=40,80)$ are small.
When $\xi=5$, 10, 15, and 20, however, the RMS velocities show good
agreement with that with $\xi=1$.
RMS power density of pressure, buoyancy and inertia are estimated
(Fig. \ref{3d_compare}). This results shows almost the same tendency as
the result of RMS velocities. Power profiles $\xi=5$, 10, 15, 20 show
agreement and those with $\xi=40$, 80 shows discrepancy with those with
$\xi=1$. These results show that the RSST is valid technique with at least
$\xi=20$ with which the Mach number is around $0.7$. 
Note that the convection pattern in our 2D and 3D cases differs
substantially. As a consequence, also the $\xi$ values for which the
validity of RSST breaks down are also different in 2D and 3D setups.
\par
\begin{figure}[htbp]
 \includegraphics[width=15cm]{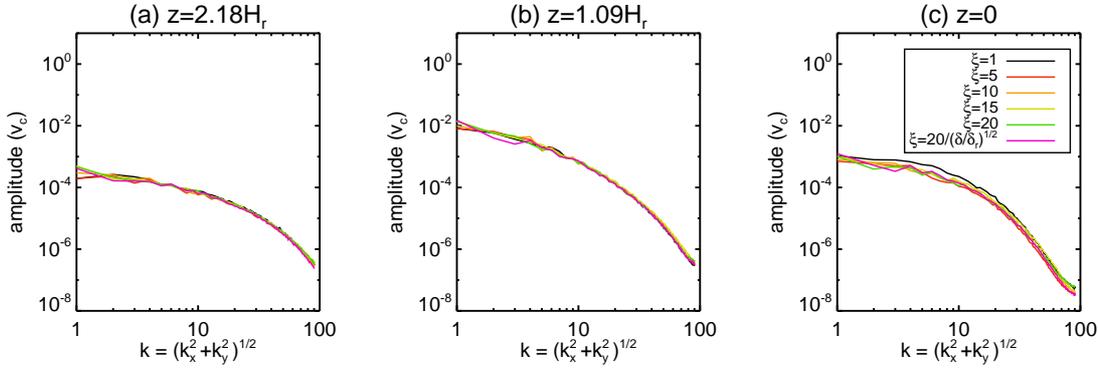}
\caption{Comparison of horizontal velocity spectra with different $\xi$.
 (a) $z=2.18H_r$ (b) $z=1.09H_r$ (c) $z=0$.  Velocities are averaged in time
 between $t=100$ and $200H_\mathrm{r}/v_\mathrm{c}$\label{compare_fft}}
\end{figure}

In Fig. \ref{compare_fft} we compare averaged spectral amplitudes for different values
of $\xi$. There is no significant difference between the spectral
amplitudes for different values of $\xi$ in the range from 1 to 20.
\par
\begin{figure}[htbp]
 \includegraphics[width=15cm]{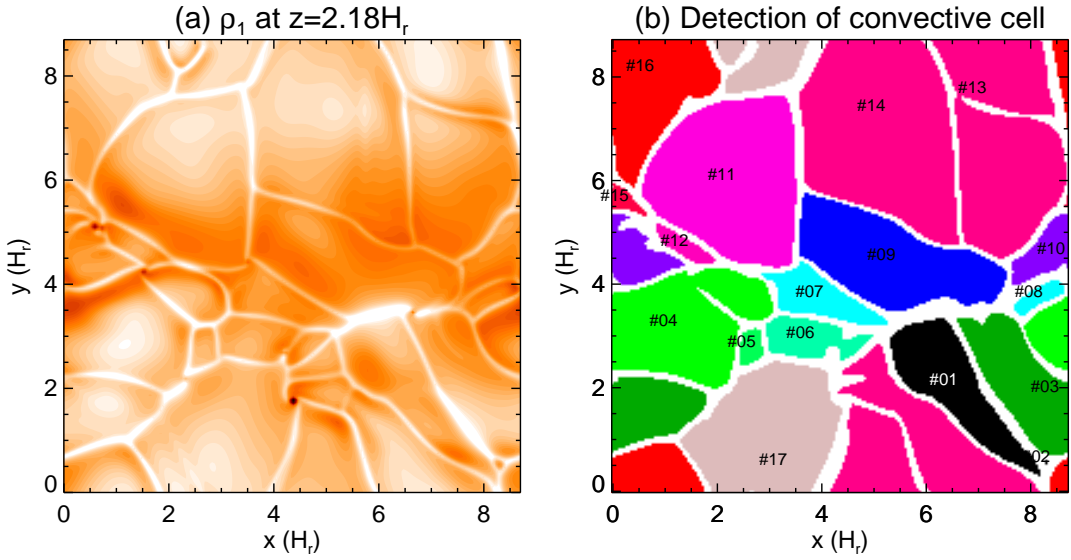}
\caption{Detection of convective cell. (a) Original contour of density
 perturbation at $z=2.18H_r$ (b) Distribution of detected convection
 cell. Color and label (\#n) show each convective cell.\label{cell_detect}}
\end{figure}
\begin{figure}[htbp]
 \includegraphics[width=15cm]{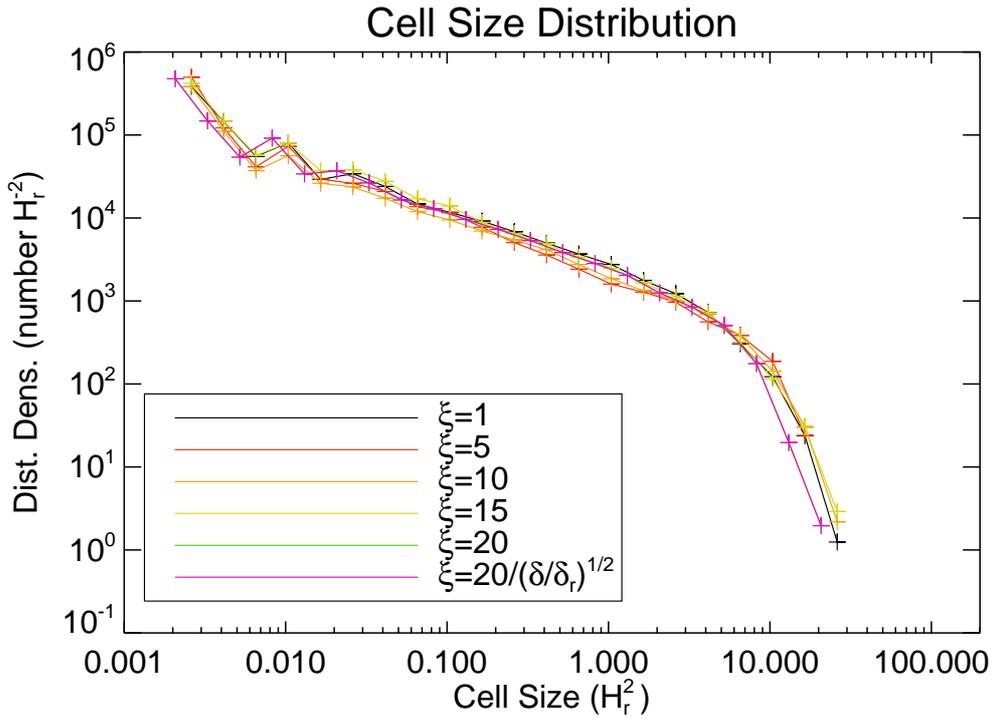}
\caption{Distribution of convective cell size with different $\xi$
 is shown. The cell size distribution is averaged in time
 between $t=100$ and $200H_\mathrm{r}/v_\mathrm{c}$.\label{cell_size_nw}}
\end{figure}

We investigate the distribution of cell size at the top boundary.
This value is significantly related to the turbulent
diffusivity and transport of angular momentum or energy.
The method to detect the cell is explained as follows. At the boundary
of the cell i.e., the region of downflow, the perturbation of density is
positive and has large value. When the
density in a region exceeds a threshold, the region is regarded as boundary of
convective cell. When a
region is surrounded by one continuous boundary region, the region is
defined as one convective cell. 
Fig. \ref{cell_detect} shows the
detected cells. Each color and each label (\#n) correspond to each
detected cell. We estimate size of all cells and compare the
distribution of cell size with different $\xi$. The results are
shown in Fig. \ref{cell_size_nw}. 
The cell size distribution follows a power law from about 0.01 to 10
$H_\mathrm{r}^2$ and there is no dependence on $\xi$ in the range from 1
to 20.
Although using this type of technique size of cells is tend to be
large with neglecting smaller cells, our conclusion is not wrong, since
all the auto detections are affected equally.
\par
With above two investigations, i.e. the Fourier analysis and study of
cell detection, we can conclude that the statistical features are not
influenced by the RSST as long as the effective Mach number
(computed with the reduced speed of sound) does not exceed values
of about 0.7, which corresponds to $\xi=20$ in our setup.\par
\begin{figure}[htbp]
 \includegraphics[width=15cm]{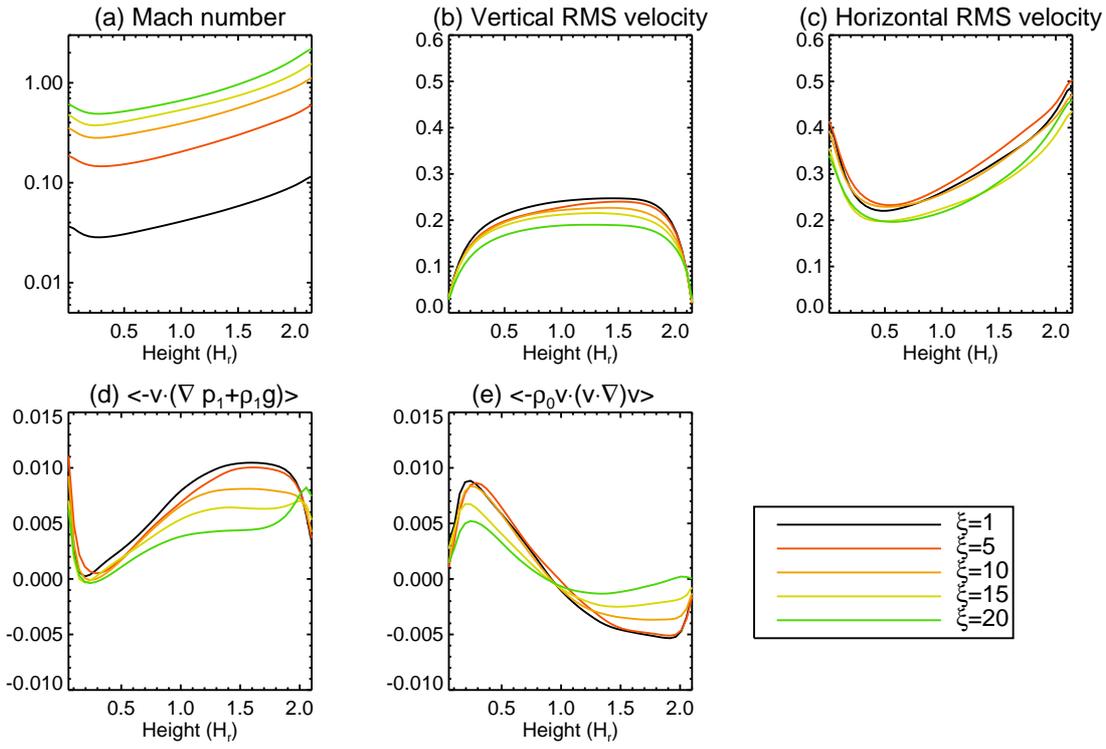}
 \caption{The results with $\delta_\mathrm{r}=1\times10^{-3}$. The
 format is the same as Fig. \ref{3d_compare}\label{lsa}}
\end{figure}

In order to confirm our criterion that the RSST is valid if the
Mach number is smaller than 0.7, we conduct calculations with
larger superadiabaticity, i.e. $\delta_\mathrm{r}=1\times10^{-3}$ (Case
14-18). If our criterion is valid, required $\xi$ with larger
superadiabaticity must decrease. The results are shown in
Fig. \ref{lsa}. Using $\delta_\mathrm{r}=1\times10^{-3}$, the calculations
with $\xi=10$, $15$, and $20$ generate supersonic convection flow near
the surface (Fig. \ref{lsa}a).
It is clear that the results with $\xi=10$,
$15$ and $20$ differ
from that with $\xi=1$ and $5$. The  calculation with
$\xi=5$ shows flow whose Mach number is around 0.6. Thus our
criterion is not violated even with different value of the superadiabaticity.\par
\begin{figure}[htbp]
 \includegraphics[width=10cm]{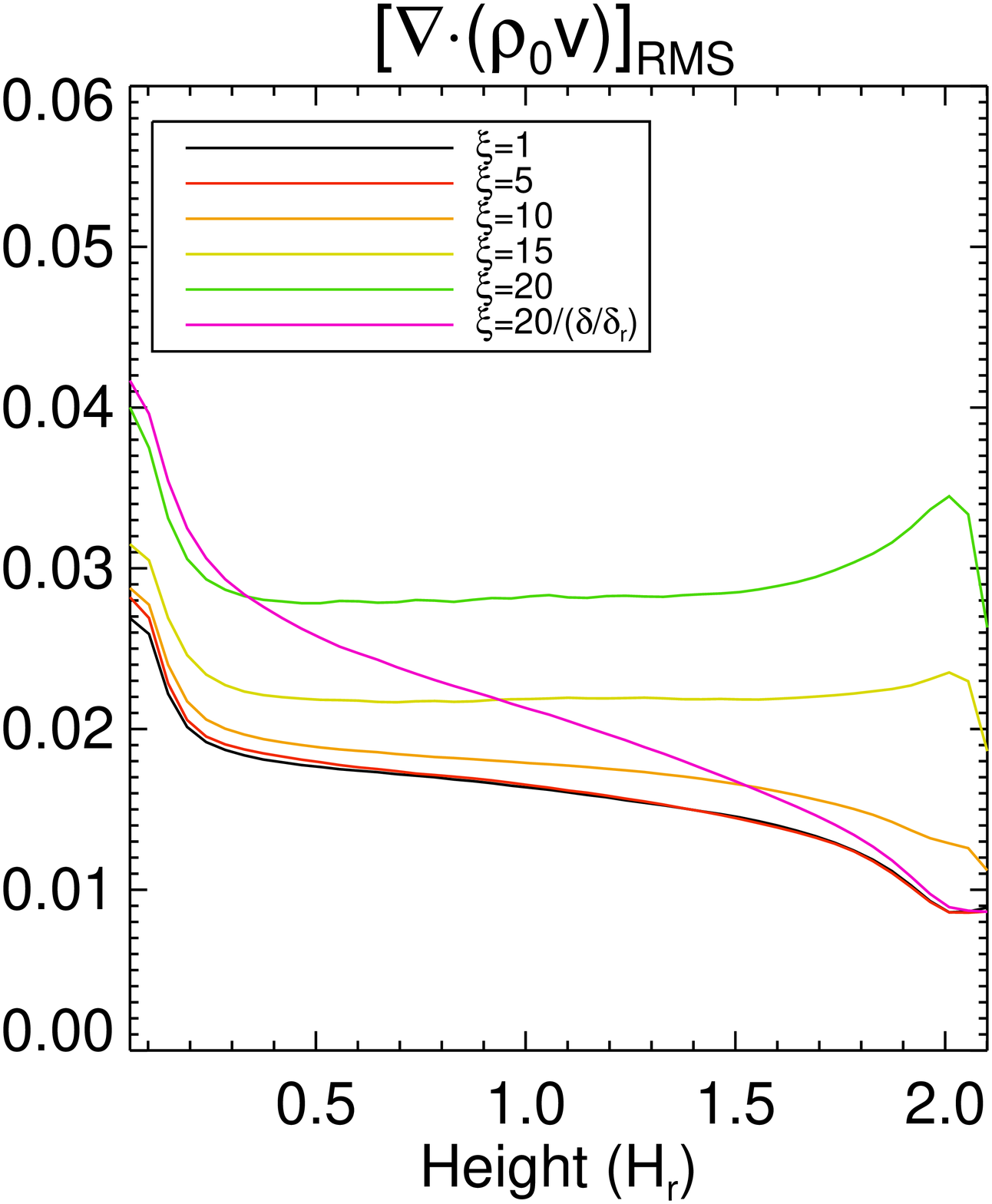}
 \caption{Dependence of
$[\nabla\cdot(\rho_0{\bf v})]_\mathrm{RMS}$ on $\xi$. In case 13, the
 value of $\xi$ is 20 at the bottom and 4.5 at the top boundary.\label{an1}}
\end{figure}

Due to our changed equation of continuity the primitive and
conservative formulation of Eq. (\ref{e:e1}) to (\ref{e:e3}) are not equivalent
anymore.  
For example, the equation of motion in conservative form is expressed as:
\begin{eqnarray}
  \frac{\partial }{\partial t}(\rho {\bf v})=
  -\nabla\cdot(\rho{\bf vv})+{\bf F},
\end{eqnarray}
where ${\bf F}$ denotes pressure gradient, gravity and Lorentz
force. With some transformations we can obtain,
\begin{eqnarray}
  {\bf v}\frac{\partial \rho}{\partial t}
+\rho\frac{\partial {\bf v}}{\partial t}
=-{\bf v}\nabla\cdot(\rho {\bf v})
-\rho({\bf v}\cdot\nabla){\bf v}+{\bf F}
\end{eqnarray}
If equation of continuity is satisfied, i.e. ($\partial \rho/\partial
t=-\nabla\cdot(\rho {\bf v})$), the primitive
form is obtained as:
\begin{eqnarray}
 \rho \frac{\partial {\bf v}}{\partial t}=\rho({\bf v}\cdot\nabla){\bf v}+{\bf
  F}.
\end{eqnarray}
However, using the RSST these two form are no longer equivalent.
We used here the primitive formulation at the expense
that energy and momentum are not strictly conserved; however,
the consistency of our results for different values $\xi$ strongly
indicates that this is not a serious problem for the setup we
considered. Alternatively we could also implement our modified
equation of continuity into a conservative formulation. This
would ensure that density, momentum and energy are strictly
conserved at the expense of a modified set of primitive
equations. Fig. \ref{an1} shows the dependence of
$[\nabla\cdot(\rho_0{\bf v})]_\mathrm{RMS}\ 
(=[\xi^2\partial \rho_1/\partial t]_\mathrm{RMS})$ on $\xi$.
Using $\xi=5$, this term is almost same as that with $\xi=1$.
Although the deviation becomes large as $\xi$ increases, it is not
proportional to $\xi^2$. \par
In the previous discussion we kept $\xi$ constant in the entire
computational domain. In the solar convection zone the Mach number
varies however substantially with depth, from $\sim 1$ in the photosphere  to
$<10^{-7}$ at the base of the convection zone, the speed of sound itself
varies from about $7\ \mathrm{km\ s^{-1}}$ in the photosphere to 
$200\ \mathrm{km\ s^{-1}}$ at the base
of the convection zone. A reduction of the speed of sound is therefore
most important in the deep convection zone, but not in the near surface
layers. This could be achieved with a depth dependent $\xi$.
Even if we use
conservative form as,
\begin{eqnarray}
  \frac{\partial \rho_1}{\partial
   t}=-\nabla\cdot\left(\frac{1}{\xi^2} \rho_0 {\bf v}\right),
   \label{non-conserve}
\end{eqnarray}
the result must not be same as the result without approximation in
statistical steady state. When the value averaged in statistical
steady state is expressed as $\langle a \rangle$, where $a$ is a
physical value, the equation of continuity becomes
\begin{eqnarray}
 0=\nabla\cdot\left(\frac{1}{\xi^2} \rho_0 \langle{\bf v}\rangle\right),
  \label{inhomogeneous}
\end{eqnarray}
with inhomogeneous $\xi$. The solution of
Eq. (\ref{inhomogeneous}) is different from the solution of original
equation of continuity in statistical steady state,
i.e. $0=\nabla\cdot(\rho_0\langle{\bf v}\rangle)$. Thus, the statistical
features such as RMS velocity are not reproduced with inhomogeneous
$\xi$.
If we use the non-conservative form of equation of motion as
Eq. (\ref{RSST}), this problem does not occur. Thus we
investigate the
nonuniformity of $\xi$ in case 13 using non-conservative form,
i.e., eq. (\ref{e:e1}).
In order to keep the Mach number
uniform in all the height, we use
$\xi=20/(\delta(z)/\delta_r)^{1/2}$, i.e. $\xi=20$ at the bottom and
$\xi=4.5$ at the top boundary.
Note that the ratio of the speed of convection to the speed of sound is
roughly estimated as $\sqrt{\delta}$. In this setting the value of
$\sqrt{\delta}$ is $1\times10^{-2}$ at the bottom and $4\times10^{-2}$
at the top boundary. The same analysis as that for
uniform $\xi$ is done (see Fig. \ref{3d_compare},
\ref{compare_fft} and \ref{an1}). There is no significant difference
between $\xi=20/(\delta(z)/\delta_r)^{1/2}$ and $\xi=1$.
Although mass is not conserved locally with
non-homogeneous $\xi$ using non-conservative form, horizontally averaged
vertical mass flux is
approximately zero in the statistically steady convection and the
conservation total mass is not significantly broken.
We conclude that an inhomogeneous $\xi$ is valid under the previously
obtained condition, i.e. the Mach number is less than 0.7.
\section{Summary and Discussion}\label{summary}
In this paper we applied RSST (see Section \ref{introduction}) to two
and three dimensional simulations of
low Mach number thermal convection and confirmed the validity of this
approach as long as the effective Mach number (computed with the
reduced speed of sound) stays below 0.7 everywhere in the domain.
The overall gain in computing efficiency that can be achieved depends
therefore on the maximum Mach number that was present in the setup of
the problem.
\par
Since the Mach number is estimated to be
$10^{-4}$ in the base of solar convection zone
\citep{2004suin.book.....S},
several thousand times longer time step can be taken using the RSST.
Therefore the RSST
and parallel computing with large number of CPUs will make it possible
to calculate large scale solar convection with high resolution in the
near future.\par
Compared to the anelastic approximation there are three major advantages
in RSST:
\begin{enumerate}
\item It can be easily implemented into any fully compressible
code (regardless of numerical scheme or grid structure) since it only
requires a minor change of the equation of continuity and leaves the
hyperbolic structure of the equations unchanged.
\item Due to the explicit treatment it does not require any additional
communication overhead, which makes it suitable for massive parallel
computations.
\item The base of the convection zone and the surface of the sun where
      the anelastic approximation is broken can be connected, using
      space dependent $\xi$.
\end{enumerate}
Overall we find that RSST is a very useful technique for studying low
Mach number flows in stellar convection zones as it substantially
alleviates stringent time step constraints without adding computational
overhead.
\begin{acknowledgements}
 The authors want to thank Dr. M. Miesch for his helpful comments.
 Numerical computations were in part carried out on Cray XT4 at Center
 for Computational Astrophysics, CfCA, of National Astronomical
 Observatory of Japan.
 We would like to acknowledge high-performance computing support provided
 by NCAR's Computational and Information Systems Laboratory, sponsored
 by the National Science Foundation.
 The National Center for Astmospheric Reseach (NCAR) is sponsored by the
 National Science Foundation. 
 This work was supported by the JSPS Institutional Program for
 Young Researcher Overseas Visits and the Research
 Fellowship from the JSPS for Young Scientists.
 We have greatly benefited from the proofreading/editing assistance
 from the GCOE program.
 We thank the referee for helpful suggestions for improvements of this paper.
\end{acknowledgements}

\begin{table}
\begin{center}
\caption{Parameters for numerical simulations. In case 13 the value of
 $\xi$ is 20 at the bottom and 4.5 at the top boudary.\label{param}}
\begin{tabular}{lcccccc}
\hline\hline
Case & dimension  & $L_x\times L_y \times L_z (H_r)$ &
$N_x \times N_y  \times N_z$ & $\delta_r$ & Re & $\xi$ \\
\hline
1 & 2 & $8.72 \times 2.18$ & $384\times 96$ & $1\times10^{-6}$ & 260 & 1 \\
2 & 2 & $8.72 \times 2.18$ & $384\times 96$ & $1\times10^{-6}$ & 260 & 10 \\
3 & 2 & $8.72 \times 2.18$ & $384\times 96$ & $1\times10^{-6}$ & 260 & 30 \\
4 & 2 & $8.72 \times 2.18$ & $384\times 96$ & $1\times10^{-6}$ & 260 & 50 \\
5 & 2 & $8.72 \times 2.18$ & $384\times 96$ & $1\times10^{-6}$ & 260 & 80 \\
6 & 3 & $8.72 \times 8.72 \times 2.18$ & $384\times 384\times 96$ & $1\times10^{-4}$ & 300 & 1 \\
7 & 3 & $8.72 \times 8.72 \times 2.18$ & $384\times 384\times 96$ & $1\times10^{-4}$ & 300 & 5 \\
8 & 3 & $8.72 \times 8.72 \times 2.18$ & $384\times 384\times 96$ & $1\times10^{-4}$ & 300 & 10\\
9 & 3 & $8.72 \times 8.72 \times 2.18$ & $384\times 384\times 96$ & $1\times10^{-4}$ & 300 & 15\\
10 & 3 & $8.72 \times 8.72 \times 2.18$ & $384\times 384\times 96$ & $1\times10^{-4}$ & 300 & 20\\
11 & 3 & $8.72 \times 8.72 \times 2.18$ & $384\times 384\times 96$ & $1\times10^{-4}$ & 300 & 40\\
12 & 3 & $8.72 \times 8.72 \times 2.18$ & $384\times 384\times 96$ & $1\times10^{-4}$ & 300 & 80\\
13 & 3 & $8.72 \times 8.72 \times 2.18$ & $384\times 384\times 96$ & $1\times10^{-4}$ & 300 & $20/(\delta/\delta_r)^{1/2}$\\
14 & 3 & $8.72 \times 8.72 \times 2.18$ & $384\times 384\times 96$ & $1\times10^{-3}$ & 300 & 1\\
15 & 3 & $8.72 \times 8.72 \times 2.18$ & $384\times 384\times 96$ & $1\times10^{-3}$ & 300 & 5\\
16 & 3 & $8.72 \times 8.72 \times 2.18$ & $384\times 384\times 96$ & $1\times10^{-3}$ & 300 & 10\\
17 & 3 & $8.72 \times 8.72 \times 2.18$ & $384\times 384\times 96$ & $1\times10^{-3}$ & 300 & 15\\
18 & 3 & $8.72 \times 8.72 \times 2.18$ & $384\times 384\times 96$ & $1\times10^{-3}$ & 300 & 20\\
\hline
\end{tabular}
\end{center}
\end{table}
\clearpage

\end{document}